\documentclass{jnmp}

\usepackage{amsmath}

\JNMPnumberwithin{equation}{section}

\begin{document}

\renewcommand{\evenhead}{S~De Lillo and M~C Salvatori}
\renewcommand{\oddhead}{On a One-Phase Stefan Problem in Nonlinear Conduction}

\thispagestyle{empty}

\FirstPageHead{9}{4}{2002}{\pageref{delillo-firstpage}--\pageref{delillo-lastpage}}{Article}

\copyrightnote{2002}{S~De Lillo and M~C Salvatori}

\Name{On a One-Phase Stefan Problem\\ in Nonlinear Conduction}
\label{delillo-firstpage}

\Author{S~DE LILLO~$^{\dag\ddag}$ and M~C SALVATORI~$^\dag$}

\Address{$^\dag$~Dipartimento di Matematica e Informatica, Universit\`{a} di Perugia, Italy\\[10pt]
$^\ddag$~Istituto Nazionale di Fisica Nucleare, sezione di Perugia, via Pascoli,\\
~~06123 Perugia, Italy}

\Date{Received December 20, 2001; Revised March 26, 2002; Accepted
April 10, 2002}

\begin{abstract}
\noindent
A one phase Stefan problem in nonlinear conduction is considered. The
problem is shown to admit a unique solution for small times. An exact
solution is obtained which is a travelling front moving with constant speed.
\end{abstract}

\section{Introduction}

One and two phase Stefan problems for the linear heat equation have been the
subject of many studies in the past [1, 2]. Indeed these problems have a
great physical relevance since they provide a mathematical model for the
processes of phase changes [3, 4].

The boundary between the two phases is a free boundary: its motion has to be
determined as part of the solution.

More recently the previous analysis was extended to nonlinear diffusion
models. In~[5,~6] exact solutions were found in parametric form for a class
of Stefan problems in nonlinear heat conduction.

Moreover one and two-phase Stefan problems for the Burgers equation were
solved in~[7,~8] and explicit travelling wave solutions were obtained.

It is the aim of this paper to formulate and solve a one-phase Stefan
problem for the nonlinear heat equation:
\begin{equation}
\frac{\vartheta _{t}}{\vartheta ^{2}}=\vartheta _{xx},\qquad \vartheta
( x,t) ,\qquad t>0,
\end{equation}
on the semiinfinite domain $x\in ( -\infty ,s(t)) $
characterized by the following set of initial and boundary data
\begin{subequations}
\begin{equation}
\vartheta ( x,0) =\vartheta _{0}( x) ,\qquad x\in
( -\infty ,b)
\end{equation}
and , $\vartheta _{0}( b) =\beta _{2}<0$, $b>0,$
\begin{gather}
\vartheta ( -\infty ,t) =\beta _{1}>0,\quad t\geq 0,\qquad \beta
_{1>}| \vartheta _{0}( x) | >| \beta _{2}|,
\qquad \vartheta _{x}( -\infty ,t) =0,\\
\vartheta ( s( t) ,t) =\beta _{2},\quad t\geq 0,\qquad \mbox{with}
\qquad s( 0) =b,\\
\vartheta _{x}( s( t) ,t) =-\dot{s}( t) .
\end{gather}
\end{subequations}
Equation (1.1) a well know mathematical model for heat conduction
in solid crystalline hydrogen for situations with one-dimensional spatial
symmetry [9]. The one-phase Stefan problem (1.1)--(1.2) is associated to a
phase change (fusion) in such a system.

In the above relations $\beta _{1}$, $\beta _{2}$ and $b$ are constants; the
unknown function $s(t)$ describes the motion of the free boundary and has to
be determined together with $\vartheta (x,t)$. Moreover, equation (1.2d) is
a condition on the flux at the free boundary, arising from heat balance
(energy) considerations.

In the following we assume $\vartheta _{0}(x)$ to be a continuously
differentiable function of its argument.

Our analysis is based on the approach followed in [7] for the solution of a
one-phase Stefan problem for the Burgers equation.

In the next Section we reduce the problem (1.1)--(1.2) to a nonlinear
integral equation for the independent variable $t$.

In Section 3 we prove existence and uniqueness of the solution for small
intervals of time; in the last Section we show that the system admits an
exact solution which travels with a constant velocity proportional to the
velocity of the free boundary.

Some details of the proof presented in the third Section are given in the
Appendix.

\section{Linearization}

In order to linearize equation (1.1) we introduce the transformation
\begin{gather}
\psi ( z,t) =\vartheta ( x,t) ,\qquad z=z(x,t) ,\qquad
z_{x}=\frac{1}{\vartheta },\qquad z_{t}=-\vartheta _{x},
\end{gather}
whose compatibility $z_{xt}=z_{tx}$is easily proved via (1.1).
Under this transformation equa\-tion~(1.1) is mapped into
\begin{equation}
\psi _{t}=\psi _{zz}
\end{equation}
on the domain $-\infty <z<\bar{z}(t)$ with $\bar{z}(t)=z(s( t) ,t) $ and $\bar{z}(0)=\bar{b}$.

(2.2) is the linear heat equation for the dependent variable $\psi
(z,t)$ with initial datum given by
\begin{subequations}
\begin{equation}
\psi ( z,0) =\psi _{0}( z_{0}) =\vartheta _{0}(x) ,
\end{equation}
where
\begin{equation}
z_{0}\equiv z_{0}( x) =\int_{-\infty }^{x}\frac{1}{\vartheta
_{0}( x') }dx'.
\end{equation}
The boundary conditions (1.2b) and (1.2c) take now the form
\begin{gather}
\psi ( -\infty ,t) =\vartheta ( -\infty ,t) =\beta _{1},\nonumber\\
\psi _{z}( -\infty ,t) =\vartheta _{x}( -\infty ,t)
\vartheta ( -\infty ,t) =0
\end{gather}
and
\begin{gather}
\psi ( \bar{z}( t) ,t) =\beta _{2}, \nonumber\\
\psi _{z}( \bar{z}( t) ,t) =-\beta _{2}\dot{s}(t) .
\end{gather}
\end{subequations}

The Stefan problem for the nonlinear equation (1.1) has then been mapped
into a~classical Stefan problem for the heat equation (2.2) with initial
datum (2.3a), characterized by the boundary conditions (2.3d) at the free
boundary.

We say $\psi (z,t),$ $\bar{z}(t)$ form a solution of the above Stefan
problem for $t<\sigma $, $0<\sigma <\infty $, when: (i) $\psi (z,t)$ is a
solution of (2.2) satisfying (2.3), it exists and is continuous
together with its derivatives for $-\infty <z<\bar{z}(t),$ $0\leq t<\sigma $;
(ii) $s(t)$ is a continuously differentiable function for $0\leq t<\sigma $.

In the following we outline a method to prove the existence and uniqueness
of the solution for small times, $t<\sigma $.

We first observe that by integrating the second relation in (2.3d) we get
\begin{subequations}
\begin{equation}
s( t) =b-\frac{1}{\beta _{2}}\int_{0}^{t}\psi _{z}( \bar{z}( t') ,t') dt',
\end{equation}
which in turn implies
\begin{equation}
\bar{z}( t) =h( t) -\frac{1}{\beta _{2}}
\int_{0}^{t}\psi _{z}( \bar{z}( t') ,t') dt',
\end{equation}
with
\begin{equation}
h(t) =\int_{0}^{s(t) }\frac{1}{\vartheta _{0}(x') }\,dx'.
\end{equation}
\end{subequations}

Next we turn our attention to the solution of (2.2). We introduce
the fundamental kernel of the heat equation
\begin{equation}
K( z-\xi ,t-\tau) =\frac{1}{2\sqrt{\pi }}\frac{1}{\sqrt{t-\tau }}
\exp \left[ -\frac{( z-\xi) ^{2}}{4( t-\tau) }
\right] ,
\end{equation}
and integrate Green's identity for the heat equation
\begin{equation}
\frac{\partial }{\partial \xi }\left( K\frac{\partial \psi }{\partial \xi }
-\psi \frac{\partial K}{\partial \xi }\right) -\frac{\partial }{\partial
\tau }( K\psi) =0,
\end{equation}
over the domain $-\infty <\xi <\bar{z}(\tau ),$ $\varepsilon <\tau
<\tau -\varepsilon $ and let $\varepsilon \rightarrow 0$. Using
$\psi (\bar{z}(\tau ),\tau )=\beta _{2}$ and $K(z-\xi ,0)=\delta
(z-\xi )$, we obtain
\begin{gather}
\psi ( z,t)  =\int_{-\infty }^{\bar{b}}K( z-\xi ,t)
\psi _{0}( \xi) d\xi-\frac{1}{\beta _{2}}\int_{0}^{t}K( z-\bar{z}( \tau)
,t-\tau) \psi _{z}( \bar{z}( \tau) ,\tau)
d\tau\nonumber\\
\phantom{( z,t)  =}{} -\beta _{2}\int_{0}^{t}K_{\xi }( \bar{z}( \tau) ,t-\tau) d\tau ,
\end{gather}
with $\bar{z}(t)$ and $h(t)$ given by (2.4b) and (2.4c)
respectively.

In the right hand side of (2.7) $\psi _{z}(\bar{z}(t),t)$ is unknown; it is
convenient to take the $z$-derivative of both sides in (2.7) and evaluate it
as $z\rightarrow z(t)^{-}$.

By putting $\nu (t) =\psi _{z}(\bar{z}(t),t)$, we obtain:
\begin{subequations}
\begin{gather}
\nu ( t)  =\left( 1+\frac{1}{2\beta _{2}}\right) ^{-1}\left[
 -\psi _{0}( \bar{b}) K( \bar{z}( t) -\bar{b},t)
+\int_{-\infty }^{\bar{b}}K( \bar{z}( t) -\xi ,t)
\psi'_{0} (\xi) d\xi   \right.\nonumber \\
\left. \phantom{\nu ( t)=} {}-\frac{1}{\beta _{2}}\int_{0}^{t}K_{z}(\bar{z}(t) -\bar{z}
( \tau) ,t-\tau) d\tau -  \beta _{2}\int_{0}^{t}K_{\tau }(\bar{z}(\tau) ,t-\tau
) d\tau \right],
\end{gather}
 with
\begin{equation}
\bar{z}( t) =h( t) -\frac{1}{\beta _{2}}\int_{0}^{t}\nu ( \tau) d\tau .
\end{equation}
\end{subequations}
Thus the solution of the Stefan problem (2.2), (2.3a), (2.3d) has
been reduced to the solution of the nonlinear integral equation (2.8a) and
(2.8b) for the independent variable~$t$.

Once the existence and uniqueness of the function $\nu ( t) $ is
established for $0\leq t<\sigma $, there follows via (2.7) the existence and
uniqueness of $\psi ( z,t) $ (and of $\vartheta ( x,t)$)
for $0\leq t<\sigma $.

\section{Contraction Mapping}

In order to analyze existence properties of $v(t)$ for $0\leq t<\sigma $, we
denote by $S_{M}$ the closed sphere $\| \nu \| <M$ in the Banach
space of functions $\nu ( t) $ continuous for $0\leq t<\sigma $
with the uniform norm $\| \nu\| =\mbox{l.u.b.}\,| \nu( t)| $.
On the sphere $S_{M}$ define the transformation
\begin{equation}
w=T\nu ,
\end{equation}
 where $T\nu $ coincides with the right hand side of (2.8a). We
first prove that $T$ is a mapping of $S_{M}$ into itself. From (2.8b) we
obtain
\begin{equation}
| \bar{z}( t)| \leq | h( t)| +\frac{1}{| \beta _{2}| }\left| \int_{0}^{t}\nu ( \tau
) d\tau \right| <\frac{M}{| \beta _{2}| }\left( 1+\frac{1}{| \beta _{2}| }\right) \sigma
\equiv B_{1}\sigma ,
\end{equation}
where (2.4c) have been used. From (2.8b) we also get
\begin{gather}
| \bar{z}( t) -\bar{z}( t')|
\leq | h( t) -h( t')| +\frac{1}{| \beta _{2}| }\left| \int_{0}^{t}\nu ( \tau)
d\tau \right|  \nonumber\\
\qquad {}<\frac{M}{| \beta _{2}| }\left( 1+\frac{1}{|\beta_{2}| }\right) | t-t'| \equiv B_{1}
|t-t'| .
\end{gather}

We now turn our attention to the right hand side of (3.1). We first note that
\begin{equation}
| K( \bar{z}( t) -\bar{b},t)| <B_{2}\sigma,
\end{equation}
where $B_{2}$ is an appropriate constant depending on $\bar{b}$.
Next we consider the integral terms in the right hand side of (3.1). We can
write:
\begin{subequations}
\begin{gather}
\left| \int_{-\infty }^{\bar{b}}K( \bar{z}( t) -\xi
,t) \psi' _{0}( \xi) d\xi \right|   \nonumber\\
\qquad {}\leq \| \psi'_{0}\| \left| \int_{-\infty }^{\bar{b}}
\frac{1}{2\sqrt{\pi }}\frac{1}{\sqrt{t}}\,\exp \left( -\frac{(
\bar{z} ( t) -\xi) ^{2}}{4t}\right) d\xi \right| \leq \frac{\|
\psi'_{0}\| }{\sqrt{\pi }}\equiv A_{1},
\end{gather}
and
\begin{gather}
\left| \beta _{2}\int_{0}^{t}K_{\tau }( \bar{z}( t) ,t-\tau )
d\tau \right|  \leq | \beta _{2}| \frac{1}{2\sqrt{\pi }}
\frac{1}{\sqrt{t}}\,\exp \left[ -\bar{z}^{2}( t) /4t\right] <|
\beta _{2}| B_{2}\sigma .
\end{gather}
Moreover we get
\begin{gather}
\left| \frac{1}{\beta _{2}}\int_{0}^{t}K_{z}( \bar{z}( t) -
\bar{z}( t') ,t-t') \nu (t') dt'\right| \nonumber\\
\qquad {}\leq \frac{M}{2| \beta _{2}| }\frac{1}{\sqrt{\pi
}}\int_{0}^{t} \frac{| \bar{z}( t) -\bar{z}( t') | }{( t-t')
^{\frac{3}{2}}}dt' < \frac{M}{| \beta _{2}| }\frac{1}{\sqrt{\pi
}}\, B_{1}\sqrt{\sigma },
\end{gather}
\end{subequations}
where (3.3) has been used.

We now use the inequality $\left( 1+\frac{1}{2| \beta _{2}| }\right) >\frac{1}{2}$
and define $M$ as $M=2A_{1}+1$. When (3.1) is used
together with (3.4) and (3.5) we get
\begin{equation}
\| w\| \leq M
\end{equation}
provided we choose $\sigma =\min ( \sigma _{1},\sigma_{2}) $
 with $\sigma _{1}: (| \beta _{2}| +| \psi_{0}( \bar{b})|) B_{2}\sigma _{1}<\frac{1}{4}$
and $\sigma _{2}:MB_{1}\sqrt{\sigma _{2}}<\frac{| \beta_{2}| \sqrt{\pi }}{4}$;
thus the mapping is closed.

Next we wish to prove that $T$ is a contraction; i.e.~given two solutions of
(3.1) with $\| \nu -\hat{\nu}\| =\delta $, it follows that
$\| t\nu -t\hat{\nu}\| =\vartheta \delta $ with $0< \vartheta < 1$.
Using (2.8b) we have for small enough $\delta $
\begin{equation}
| \bar{z}( t) -\hat{\bar{z}}(t')| \leq \frac{1}{| \beta _{2}|
}\delta t\left| \int_{\hat{s}( t) }^{s( t)
}\frac{1}{\vartheta_{0}( x') } dx'\right| \leq \frac{1}{|\beta
_{2}| }\, \delta t\left( 1+\frac{1}{| \beta _{2}| } \right) \equiv
B_{3}\delta t,
\end{equation}
where (2.4c) has also been used.

Similar estimates hold for $\dot {\bar{z}}(t)$, which will be useful
in the following. From (2.8b) we see that $\dot{\bar{z}}(t)$ is
bounded
\begin{subequations}
\begin{equation}
| \dot{\bar{z}}(t)| \leq | \dot{h}( t)
| +\frac{1}{| \beta _{2}| }| \nu ( t)
| \leq \frac{M}{| \beta _{2}| }\left( 1+\frac{1}{|\beta _{2}| }\right) \equiv B_{3}M,
\end{equation}
moreover it is
\begin{equation}
| \dot{\bar{z}}( t) -\dot{\hat{z}}(t')| \leq \frac{3}{| \beta
_{2}| }\, \delta .
\end{equation}
\end{subequations}

>From (3.1) we now write
\begin{subequations}
\begin{equation}
w-\hat{w}=\left( 1+\frac{1}{2\beta _{2}}\right) ^{-1}\sum_{i=1}^{4}H_{i}
\end{equation}
with
\begin{gather}
H_{1=}\psi _{0}( \bar{b}) \left[ K( \hat{\bar{z}}( t) -\bar{b},t) -
K( \bar{z}( t) -\bar{b},t) \right] ,\\
H_{2}=\int_{-\infty }^{\bar{b}}\psi'_{0}( \xi) \left[
K( \bar{z}(t) -\xi ,t) -K( \hat{\bar{z}}(t) -\xi ,t) \right] d\xi , \\
H_{3} =\frac{1}{\beta _{2}}\int_{0}^{t}\left[  K_{z}(
\hat{\bar{z}}(t) -\hat{\bar{z}}( t') ,t-t') \nu (t')
-K_{z}(\bar{z}(t) -\bar{z}(t')
,t-t') \hat{\nu}(t') \right]dt' ,\\
H_{4}=\beta _{2}\int_{0}^{t}\left[ K_{t'}( \hat{\bar{z}}( t) ,t-t') -K_{t'}(\bar{z}(t) ,
t-t') \right] dt'.
\end{gather}
\end{subequations}
First we estimate $H_{1}$. We use the mean value theorem together with (3.7)
and (3.2) in the right hand side of (3.9b); we get
\begin{equation}
| H_{1}| \leq \| \psi _{0}\| \frac{1}{4\sqrt{\pi t}}\,
B_{1}B_{3}\delta t<\frac{\| \psi _{0}\| }{4\sqrt{\pi t}} \,
B_{1}B_{3}\sqrt{\sigma }\,\delta \equiv B_{4}\sqrt{\sigma
}\,\delta .
\end{equation}

The estimate of $H_{2}$ in (3.9c) is obtained by writing
\begin{subequations}
\begin{equation}
| H_{2}| \leq \| \psi'_{0}\| \frac{1}{\sqrt{\pi }}
\left| \int_{\bar{y}}^{\hat{\bar{y}}}e^{-y^{2}}dy\right| ,
\end{equation}
with $\bar{y}=\frac{ \bar{z}(t) -\bar{b}}
{2\sqrt{t}}$ and $\hat{\bar{y}}=\frac{ \hat{\bar{z}}( t) -\bar{b} }{2\sqrt{t}}$; we then
obtain
\begin{equation}
|H_{2}| \leq \| \psi'_{0}\| \frac{1}{2\sqrt{\pi }}\, | \bar{z}( t)
-\hat{\bar{z}}(t)| <\frac{\| \psi'_{0}\| }{2\sqrt{\pi }}\,
B_{3}\sqrt{\sigma }\,\delta \equiv B_{5}\sqrt{\sigma }\,\delta ,
\end{equation}
\end{subequations}
 where (3.7) has been used.

The estimate of $H_{3}$ is somewhat more cumbersome; a detailed analysis is
given in the Appendix (see (A.1)--(A.6)).

There obtains
\begin{equation}
| H_{3}| <B_{6}\sqrt{\sigma }\,\delta
\end{equation}
where $B_{6}$ is defined in (A.7).

We finally turn our attention to the estimate of $H_{4}$ in (3.9e). When
(2.5) is used, we get from the integral in the right hand side of (3.9e)
\begin{gather}
|H_{4}|  \leq \frac{| \beta _{2}| }{2\sqrt{\pi t}}\left|\exp (
-\hat{\bar{z}}{}^{2}( t) /4t) -\exp ( -\hat{\bar{z}}{}^{2}( t)
/4t)\right|\! <\frac{| \beta _{2}| }{4\sqrt{\pi
}}\,B_{1}B_{3}\sqrt{\sigma } \delta \equiv B_{7}\sqrt{\sigma }\,
\delta ,
\end{gather}
where use of the mean value theorem together with (3.2) and (3.7)
has been made.

>From (3.9a) we now write
\[
| w-\bar{w}| \leq \frac{2| \beta _{2}| }{1+2| \beta _{2}|}\sum_{i=1}^{4}| H_{i}| ,
\]
which in turn implies, when we combine together the estimates
(3.10)--(3.13):
\begin{equation}
\frac{\| w-\bar{w}\| }{\delta }<\sqrt{\sigma }
\sum_{i=4}^{7}B_{i}\equiv \sqrt{\sigma }\,B_{8};
\end{equation}
thus we conclude that if $\sigma $ satisfies $\sigma <\min (\sigma_{1},
\sigma_{2},\sigma_{3}) $ where
\begin{equation}
\sqrt{\sigma }\,B_{8}<\sigma
\end{equation}
it follows that $T$ is a contraction operator on $S_{M}$, which
admits a unique fixed point $v=Tv$ in $S_{M}$ for $0\leq t<\sigma $.

We have then proven the existence and uniqueness of the solution of the
integral equation (2.8a) for a small interval of time.

\section{A Particular Solution}

We now turn our attention to a particular solution of the Stefan problem
(1.1), (1.2). Namely, we consider a moving front solution of equation~(2.2)
\begin{subequations}
\begin{equation}
\psi( z,t) =\beta _{1}\left( 1-e^{-V( z-Vt) }\right) ,
\end{equation}
with
\begin{equation}
V<0,
\end{equation}
\end{subequations}
which is travelling to the left with constant speed $V$ and is
compatible with the boundary conditions (2.3c). We now impose on (4.1a) the
Stefan boundary conditions (2.3d): the first one implies
\begin{subequations}
\begin{equation}
\bar{z}=\bar{b}+Vt,
\end{equation}
 which in turn gives
\begin{equation}
\dot{\bar{z}}=V=-\frac{1}{b}\ln \left( 1+\frac{| \beta_{2}| }{\beta _{1}}\right) .
\end{equation}
\end{subequations}

The boundary $\bar{z}( t) $ and the front solution (4.1a) are
then both moving to the left with the same constant velocity.

When we next use the second boundary condition (2.3d), keeping into account
the first one, we obtain
\begin{equation}
\dot{s}( t) =\frac{( \beta _{2}-\beta _{1}) }{\beta _{2}}\, V,
\end{equation}
which shows that the moving boundary $s(t) $ of the
Stefan problem (1.1)--(1.2) is moving to the left with constant speed
$\dot{s}=\alpha V$, $\alpha =\frac{\beta _{2}\beta _{1}}{\beta _{2}}>1$.

Finally, the solution of the one-phase Stefan problem for the nonlinear heat
equation~(1.1) is given by
\begin{equation}
\vartheta ( x,t) =\left( \frac{\partial z}{\partial x}\right)^{-1},
\end{equation}
where, in virtue of (2.1), $z(x,t)$ solves
\begin{equation}
x=\int_{0}^{z}\psi ( z',t) dz',
\end{equation}
with $\psi (z,t)$ given by (4.1a) and the speed $V$ specified by
(4.2b).

We emphasize that the above solution is a very special solution of the
Stefan problem~(1.1), (1.2). Indeed, it corresponds to particular case
when the nonlinear integral equation (2.8) reduces to a linear integral
equation of Volterra type in $t$, as implied by substituting back (4.2a)
into (2.8a).

\renewcommand{\theequation}{A.\arabic{equation}}
\setcounter{equation}{0}

\section*{Appendix}

In order to estimate $H_{3}$, starting from (3.9d) we write
\begin{gather}
H_{3}=\frac{1}{\beta _{2}}\left[
-\int_{0}^{t}dt'\bar{\nu}( t') \frac{\hat{\bar{z}}( t) -
\hat{\bar{z}}( t') }{t-t'}K( \hat{\bar{z}}( t) -\hat{\bar{z}}( t') ,t-t')  \right.\nonumber\\
\left.\phantom{H_{3}=}{}+\int_{0}^{t}dt'\nu ( t') \frac{\bar{z}(t) -\bar{z}( t') }
{t-t'}K( \bar{z}( t) -\bar{z}( t') ,t-t') \right],
\end{gather}
with $K(z-\xi ,t-\tau )$ given by (2.5).

Next, we put
\begin{subequations}
\begin{gather}
H_{3}=\frac{1}{\beta _{2}}( J_{1}+J_{2}+J_{3}),\\
J_{1}=-\int_{0}^{t}dt'\left\{ ( \hat{\nu}( t') -\nu ( t')) \frac{\hat{\bar{z}}(t) -
\hat{\bar{z}}( t') }{t-t'}\,K( \hat{\bar{z}}(t) -\hat{\bar{z}}( t'),t-t') \right\} ,\\
J_{2} =-\int_{0}^{t}dt'\left\{ \nu ( t') \left[
\frac{\hat{\bar{z}}(t) -\hat{\bar{z}}( t') }{t-t'}-\frac{\bar{z}(t) -\bar{z}( t') }{t-t'}\right]
K(\hat{\bar{z}}( t) -\hat{\bar{z}}( t'),t-t')\right\} ,\\
J_{3} =-\int_{0}^{t}dt' \Bigg\{ \nu ( t') \frac{\bar{z}(t)
-\bar{z}( t') }{t-t'} \,K( \bar{z}( t) -\bar{z}( t')
,t-t')  \nonumber\\
\phantom{J_{3} =}{}\times\left[ 1-\exp \left\{ -\frac{( \hat{\bar{z}}(t) -\hat{\bar{z}}( t'))
^{2}-( \bar{z}( t) -\bar{z}( t'))
^{2}}{4( t-t') }\right\} \right]\Bigg\}.
\end{gather}
\end{subequations}

By using (3.3) we estimate $J_{1}$:
\begin{equation}
| J_{1}| \leq \frac{1}{\sqrt{2\pi }}\int_{0}^{t}| \hat{\nu}
( t') -\nu ( t')| \left|
\frac{\hat{\bar{z}}(t) -\hat{\bar{z}( t') }}{t-t'}\right| \frac{dt'}
{\sqrt{t-t'}}\leq \left( \frac{B_{1}}{\sqrt{\pi }}\sqrt{\sigma }
\right) \delta .
\end{equation}

The estimate of $J_{2}$ uses (3.8b) and the mean value theorem:
\begin{gather}
| J_{2}|  \leq \frac{1}{\sqrt{2\pi }}\int_{0}^{t}| \nu( t')|
\left| \frac{\hat{\bar{z}}( t) -\hat{\bar{z}}( t')}{t-t'}
-\frac{\bar{z}(t) -\bar{z}(t') }{t-t'}\right|
\frac{dt'}{\sqrt{t-t'}}\,
\delta   \nonumber \\
\phantom{|J_{2}|}{}\leq \frac{M}{\sqrt{\pi }}\int_{0}^{\sigma }| \hat{\bar{z}}( \vartheta) -
\dot{\bar{z}}(\vartheta)| \frac{dt'}{\sqrt{t-t'}}\leq
\left( 3\frac{M}{\sqrt{\pi }}\frac{1}{| \beta _{2}| }\sqrt{\sigma }\right) \delta .
\end{gather}

Finally, for the estimate of $J_{3}$, we call
\begin{subequations}
\begin{gather}
Q =-\frac{( \hat{\bar{z}}(t) -\hat{\bar{z}}( t')) ^{2}-
(\bar{z}(t) -\bar{z}( t')) ^{2}}{4( t-t') }  \nonumber \\
\phantom{Q}{}=\frac{[( \bar{z}( t) -\hat{\bar{z}}( t')) -( \bar{z}( t) -{\bar{z}}( t'))]
[( \bar{z}( t) -\bar{z}( t')) -(\hat{\bar{z}}(t) -\hat{\bar{z}}( t'))] }{4( t-t') }.
\end{gather}

Then from (3.3) and (3.7) we have
\[
|Q| <\frac{1}{4| t-t'| }\,4B_{1}B_{3}|t-t'| \sigma \delta ,
\]
which gives
\begin{gather}
| Q| <B_{1}B_{3}\sigma \delta .
\end{gather}

We also note that we can estimate $Q$ via (3.3):
\begin{gather}
|Q|  <\frac{1}{4| t-t'| }\left[(\hat{\bar{z}}(t) -\bar{z}( t'))^{2}+(\bar{z}(t) -\bar{z}(t'))^{2}\right]
\nonumber\\
\phantom{|Q|}{}\leq \frac{1}{2}B_{1}^{2}| t-t'| <\frac{1}{2}B_{1}^{2}\sigma <B_{1}^{2},\qquad
 \frac{\sigma }{2}<1.
\end{gather}
\end{subequations}

Using $\left| 1-e^{-Q}\right| \leq | Q| e^{|Q| }$ and
(3.3) we estimate $J_{3}$:
\begin{gather}
| J_{3}|  \leq \frac{1}{2\sqrt{\pi }}\int_{0}^{\sigma }|\nu ( t')|
\left| \frac{\bar{z}( t) -\bar{z}( t') }{t-t'}\right| \frac{1}{\sqrt{t-t'}}
\left| 1-e^{-Q}\right| dt'  \nonumber \\
\phantom{|J_{3}|}{}<\frac{M}{\sqrt{\pi
}}\,B_{1}^{2}B_{3}e^{B_{1}^{2}}\sigma ^{3/2}\delta <\left( \frac{|
\beta _{2}| }{4}B_{1}B_{3}e^{B_{1}^{2}}\sqrt{\sigma }\right)
\delta ,
\end{gather}
where the definition of $\sigma $ following (3.6) has also been
used.

Combining the estimates of $J_{1}$, $J_{2}$ and $J_{3}$, we have from (A2.a)
\begin{equation}
| H_{3}| =\frac{1}{| \beta _{2}| }\left( \frac{B_{1}}{\sqrt{\pi
}}+\frac{3M}{\sqrt{\pi }}\frac{1} {| \beta _{2}| }+\frac{| \beta
_{2}| }{4}\, B_{1}B_{3}e^{B_{1}^{2}}\right) \sqrt{\sigma }\,
\delta \equiv B_{6}\sqrt{\sigma }\,\delta .
\end{equation}

\label{delillo-lastpage}

\end{document}